\documentclass{elsart}
\usepackage{amsfonts}
\usepackage{amssymb}
\usepackage{amsbsy}

\begin{document}

\begin{frontmatter}

\title{Lie-type transformations and effective 
Hamiltonians in nonlinear quantum optics: 
applications to multilevel systems}

\author{A. B. Klimov, A. Navarro}
\address{Departamento de F\'{\i}sica, 
Universidad de Guadalajara,  
Revoluci\'on 1500, 
44420~Guadalajara, Jalisco, 
Mexico} 

\author{L. L. S\'anchez-Soto} 
\address{Departamento de \'Optica, 
Facultad de Ciencias F\'{\i}sicas, 
Universidad Complutense, 
28040 Madrid, Spain}

\maketitle

\begin{abstract}
We reelaborate on a general method for diagonalizing 
a wide class of  nonlinear Hamiltonians describing 
different quantum optical models. This method makes 
use of  a nonlinear deformation of the usual su(2) algebra 
and  when some physical parameter, dictated by 
the  particular model under consideration, becomes
small, it gives a diagonal effective Hamiltonian 
that describes correctly the dynamics for arbitrary 
states and long times. We extend the technique to 
$N$-level atomic systems interacting with quantum 
fields, finding the corresponding effective Hamiltonians
when the condition of $k$-photon resonance is fulfilled. 

\textit{PACS number(s): 03.65.Bz,  03.65.Ca, 
42.50.Hz, 42.50.Fx}
\end{abstract}

\end{frontmatter}

\section{Introduction}

Many nonlinear quantum optical processes, such as $k$-th 
harmonic generation, $k$-wave mixing or generalized Dicke 
models, can be described in terms of operators from a nonlinear 
or deformed su(2) algebra~\cite{1,2,3,4,5,6,7,8}. The emergence
of this structure allows us to handle the problem in a close 
analogy with the usual treatment for an angular momentum:
in consequence, we get a decomposition of the Hilbert space into 
direct  sums of invariant subspaces and, therefore, the dynamical 
problem generated by the corresponding Hamiltonian can be 
reduced to the diagonalization of a finite-dimensional matrix. 
In spite of this considerable achievement, convenient analytic 
expressions for the eigenvalues and eigenstates are unknown, 
although several  approximations have been devised relying on 
the specific features of the particular problem under study. 

Recently~\cite{9}, the method of Lie transformations (see,  e.g. 
Ref.~{10} and references therein) has been applied to the 
approximate diagonalization of nonlinear optical Hamiltonians 
that can be recast in terms of the generators of polynomial 
deformed algebras. In this paper we further pursue this topic 
by applying the method to the relevant case of multilevel atomic 
systems interacting with single-mode quantized fields. 
Consequently, we find the corresponding effective Hamiltonians 
when the condition of $k$-photon resonance is fulfilled. 

The contents of this paper are organized as follows: after a 
brief outline of the general method of small rotations developed 
in  Ref.~\cite{9}, we show how it works for multilevel systems and,
finally, we give explicit examples for the cases of two- and 
three-photon resonances.

\section{Nonlinear su(2) dynamics and effective Hamiltonians}

Let us assume that the interaction Hamiltonian  of the model under
consideration can be written as 
\begin{equation}
H_{{\rm int}} = \Delta \ X_3 + gV  ,  \label{Hint}
\end{equation}
where 
\begin{equation}
V = X_+ + X_- .  \label{Vint}
\end{equation}
Here $g$ is a coupling constant that can be always chosen as
real, $\Delta $ is a parameter usually representing the 
detuning between frequencies of different subsystems
(although it is not necessary), and the operators $X_\pm$ and 
$X_3$ maintain the first commutation relation of the su(2) algebra 
\begin{equation}
[X_3, X_\pm ] = \pm X_\pm ,  \label{Xpm}
\end{equation}
while the second one is modified in the following way 
\begin{equation}
[X_+, X_- ] = P(X_3) ,  \label{P}
\end{equation}
where $P(X_3)$ is an arbitrary polynomial function of the 
diagonal operator $X_3$ with coefficients that may depend on 
some integrals of motion $N_j$. 

Now suppose that for some physical reason (depending on the
particular model under consideration) the condition 
\begin{equation}
\varepsilon =\frac{g}{\Delta }\ll 1
\end{equation}
is fulfilled. Then, it is clear that the Hamiltonian (\ref{Hint}) is 
\textit{almost} diagonal in the basis of the eigenstates of $X_3$. 
In fact, a standard perturbation analysis immediately shows 
that the first-order corrections introduced by the nondiagonal 
part $g V$ to the eigenvalues of $X_3$ vanish and those of 
second order are proportional to $g/\Delta \ll 1$. To proceed
further we apply to (\ref{Hint}) the following unitary 
transformation  (which, in fact, is a \textit{small} nonlinear rotation) 
\begin{equation}
U=\exp ( \varepsilon T ) ,
\label{Umain}
\end{equation}
with 
\begin{equation}
T = X_+ - X_- , 
\label{Tmain}
\end{equation}
in such a way that 
\begin{equation}  \label{Heff}
H_{{\rm eff}} = UH_{{\rm int}}
U^{\dagger }.  \label{H1}
\end{equation}
Using the standard expansion 
\begin{equation}
e^A B e^{-A} = B + [A, B] +
\frac{1}{2!} [A, [A, B]] +\ldots ,
\end{equation}
we obtain 
\begin{equation}
H_{{\rm eff}} = \Delta \ X_3 + g 
\sum_{k=1}^\infty
\varepsilon^k \frac{k}{(k+1)!}{\rm ad}_{T}^k (V),
\end{equation}
where ${\rm ad}_{T}$ is the adjoint operator defined as 
${\rm ad}_{T}(V)=[T,V].$  In other words, the effective 
Hamiltonian takes the form 
\begin{eqnarray}
H_{{\rm eff}} & = & \Delta \ X_3 - 
\varepsilon g \ \nabla \Phi (X_3) +
\varepsilon^2 \frac{2g}{3}   \ 
[ X_+ \nabla^2 \Phi (X_3) + 
\nabla^2 \Phi (X_3) X_- ]  \nonumber \\
& - & 
\varepsilon^3 \frac{ g}{4} \ 
\{ X_+^2 \nabla^3 \Phi (X_3) +
\nabla^3 \Phi (X_3) X_-^2 +
2 \nabla [ \Phi (X_3) \nabla^2 \Phi (X_3 - 1) ] \}
\nonumber \\
& + &  \mathcal{O} (\varepsilon^4). 
\end{eqnarray}
where $\Phi (X_3) = X_+ X_-$ is a structural function such that 
\begin{equation}
P(X_3) = - \nabla \Phi (X_3) = 
\Phi (X_3) - \Phi (X_3+1).
\end{equation}
By keeping terms up to order $\varepsilon$ we finally get 
\begin{equation}
H_{{\rm eff}}=\Delta \ X_3 + \frac{g^2}{\Delta}P(X_3),
\label{H1eff}
\end{equation}
and the resulting effective Hamiltonian is diagonal in the 
basis of eigenstates of  $X_3$.

Furthermore, the higher-order contributions have always 
the form $X_+^k f(X_3) + f(X_3) X_-^k + g(X_3)$. This 
makes the procedure of removing the off-diagonal terms 
somehow trivial at each step, in the sense that it is always 
obvious the appropriate transformation that should be applied. 
For example, to eliminate terms of the form 
\begin{equation}
\varepsilon^\ell 
[ X_+^k f(X_3) + f(X_3) X_-^k ] , 
\end{equation}
it suffices with applying the transformation 
\begin{equation}
\exp (\varepsilon^{\ell+1}T_k/k ) ,  \label{Tp}
\end{equation}
with $T_k = X_+^k f(X_3) - f(X_3) X_-^k$, since the first 
commutator of $\Delta \ X_{3}$ with $T_k$ cancels the
corresponding term in the Hamiltonian.

With this approach, the evolution (as well as the spectral) 
problem is completely solved.  The method has the virtue 
of generality, since it is valid for any model whose Hamiltonian
could be written down in terms of the generators of an 
arbitrary polynomial deformation of su(2), such as, 
for example, three-wave mixing, Dicke model or $k$-th harmonic 
generation [9]. Because the transformed Hamiltonian is an 
expansion in the small parameter $\varepsilon $, it is clear 
that its eigenvalues coincide with those obtained using the 
standard perturbation theory. The advantage of our method 
lies in obtaining \textit{diagonal effective Hamiltonians} 
in an operator form, which avoids the necessity of calculating 
cumbersome series of matrix elements.

This technique also provides corrections to the eigenstates 
of the Hamiltonian~(\ref{Hint}). Indeed, from~(\ref{H1}) it is 
easy to realize that the eigenstates of the interaction 
Hamiltonian~(\ref{Hint}) can be approximated as 
\begin{equation}
|\Psi _m \rangle = U^{\dagger} |m\rangle ,
\end{equation}
where $ |m \rangle $ is an eigenstate of $X_3$ 
and $U$ is the corresponding small rotation. Since 
the rotation operators and $|m\rangle $ are time independent, 
the operator $U$ can be applied to $|m\rangle $ in the form 
of an expansion in $\varepsilon $. For example, the 
eigenstate $|\Psi_m \rangle $ up to order $\varepsilon^2$ 
takes the form 
\begin{eqnarray}
|\Psi_m \rangle  & = &  | m \rangle - 
\varepsilon ( X_+ - X_- ) | m \rangle \nonumber \\
& + & 
\frac{\varepsilon^2}{2} 	\{ (X_+^2 + X_-^2 )  - 
[ \Phi (X_3 ) + \Phi ( X_3 + 1) ] \} | m \rangle .
\end{eqnarray}
This representation is especially advantageous if we construct the 
space of states of the model as a representation space of the 
deformed su(2) algebra using the raising operator $X_+$~\cite{1,2,3}: 
\begin{equation}
| m \rangle \propto X_+^m | 0 \rangle ,
\end{equation}
where $| 0 \rangle $ is the lowest weight vector fulfilling 
the standard condition $X_- | 0 \rangle =0$.

\section{Nonlinear su(N) dynamics and effective Hamiltonians}

The method of approximate diagonalization embodied in 
Eqs.~(\ref{Hint}), (\ref{Umain}), and (\ref{H1eff}) can be 
applied  also to systems with a more complicated algebraic 
structure.

In what follows we consider Hamiltonians which can be 
represented in terms of the su(N) algebra. This algebraic 
structure naturally arises when describing atomic systems 
with $N$ relevant energy levels.

The Hamiltonian governing the evolution of a collection of 
$A$ identical $N$-level atoms (for definiteness, we consider 
the case of a cascade configuration, such that $E_i < E_j$ for 
$i<j$) interacting with a single-mode quantum field of frequency 
$\omega_f$ has te form  (in units $\hbar = 1$)
\begin{equation}
H = \omega_f a^\dagger a
+ \sum_{j=1}^N \omega_j S^{jj} + 
\sum_{j=1}^{N-1}g_j (aS_+^{j j+1}+
a^\dagger S_-^{j j+1}),  \label{H}
\end{equation}
where $S^{ii}$ ($i=1, \ldots, N$) are the population operators of 
the $i$-th level of energy $\omega_i$, and $S_+^{ij} = S^{ij},$ 
$S_-^{ij}= ( S_+^{ij})^\dagger $ ($i<j$) describe transitions 
between levels  $i$ and $j$. The operators $S^{ij}$ form the  
algebra u(N) and satisfy the commutation relations 
\begin{equation}
[ S^{ij}, S^{km}] = 
\delta_{im} S^{jk} + \delta _{jk}S^{im}. 
\end{equation}
Due to the integral of motion 
\begin{equation}
\sum_{j=1}^N S^{jj}=A,  \label{SA}
\end{equation}
one can introduce inversion operators $(1\leq j \leq N-1)$
\begin{equation}
S_z^{j j+i} = \frac{1}{2}
( S^{j+1 j+1}- S^{jj} ) . 
\end{equation}
In this way the operators ($S_\pm^{ij}, S_z^{j j+1}$) form 
the algebra su(N). Apart from~(\ref{SA}), the 
Hamiltonian~ (\ref{H}) possesses the following 
integral of motion 
\begin{equation}
\hat{N}= a^{\dagger }a + 
\sum_{j=1}^{N-1} \mu_j S_z^{jj+1},  \label{IM}
\end{equation}
with $\mu _{j}=j(N-j)$, that represent the total number of excitations
in the system. Let us introduce detunings between energy levels of
the atomic system according to 
\begin{equation}
\Delta_j = \omega_j - \omega_1 - ( j-1) \omega_f  , 
\end{equation}
with the trivial condition that $\Delta_1 = 0$. Moreover, let
us assume that $\Delta_j$ satisfy the following resonant condition 
\begin{equation}
\Delta_N = 0,  \label{rc}
\end{equation}
which means that the field is in a $(N-1)$-photon resonance 
with the  atomic system: $\omega_N - \omega_1 = (N-1) \omega_f$. 
Then,  the Hamiltonian (\ref{H}) can be rewritten as
\begin{equation}
H = H_0 +  H_{{\rm int}} ,
\end{equation}
with 
\begin{eqnarray}
H_0 & = &  \omega_f  \hat{N} + \omega A \nonumber \\
& &  \nonumber  \\
H_{{\rm int}} & = & h_0 +V , 
\end{eqnarray}
where
\begin{eqnarray}
\omega & = & \frac{1}{2} (\omega_N + \omega_1) , \nonumber \\
h_{0} & = & \sum_{j=1}^N \Delta_j S^{jj} , \nonumber \\
V & = & \sum_{j=1}^{N-1}g_j (a S_+^{j j+1} +
a^\dagger S_-^{j j+1}) . 
\end{eqnarray}
One can observe that the operators $(1 \leq  i < j \leq N)$
\begin{eqnarray}
& X_+^{ij}=aS_{+}^{ij}, \quad 
X_-^{ij}=a^\dagger S_-^{ij}, & \nonumber \\
& & \nonumber \\
& X^{kk}=S^{kk} , &  
\end{eqnarray}
form a polynomial deformation of the algebra u(N), since
they satisfy the following commutation relations 
\begin{eqnarray}
& \ [X^{ii}, X^{kk}]=0 , \quad 
[X_+^{ij}, X_+^{ik}]=0 , \quad
[X_-^{ij}, X_-^{ik}] = 0 , &  \nonumber \\
& \ [X^{kk}, X_\pm^{ij}] = 
\pm X_\pm^{ij} ( \delta _{kj}-\delta_{ik}) , \nonumber \\
& [X_+^{ij}, X_-^{ij}] = P(X^{ii},X^{kk}), & 
\end{eqnarray}
where $P(X^{ii},X^{kk})$ is a polynomial function of 
the diagonal operators $X^{ii}$ and the integral of motion 
$\hat{N}$.

According to the general scheme (\ref{Umain})-(\ref{H1eff}) 
we introduce the following transformation 
\begin{equation}
U_{1}=\exp (T_1) ,
\end{equation}
with
\begin{equation}
T_1=\sum_{j=1}^{N-1} \alpha_j^{(1)}
(aS_+^{jj+1} - a^\dagger S_-^{j j+1}),  
\label{T1}
\end{equation}
where the parameters $\alpha_j^{(1)}$ are
\begin{equation}
\alpha_j^{(1)}=\frac{g_j}
{\Delta_{j+1}-\Delta _j}  
\label{alfa}
\end{equation}
and are assumed to be small numbers, $\alpha_j^{(1)}\ll 1$, 
which means that the atomic transitions are far from 
the one-photon resonance ($\Delta_{j+1}-\Delta_j =
\omega_{j+1} - \omega_j - \omega_f \gg g_j$). 
Since 
\begin{equation}
[ T_1 , h_0 ] = -V, 
\end{equation}
all one-photon transitions are eliminated by (\ref{T1})
and the transformed Hamiltonian $H_{{\rm eff}}^{(1)}=
U_1 H_{{\rm int}} U_1^\dagger$ takes the form 
\begin{eqnarray}
H_{{\rm eff}}^{(1)}& = & 
h_0 + h_{{\rm diag}}+h_{{\rm nondiag}} \nonumber \\
& + &  
\sum_{k=1}^{N-2} \frac{k}{(k+1)!}
\sum_{j=1}^{N-k-1}\psi_j^{(k+1)}
(a^{k+1}S_+^{j j+k+1} + {a^\dagger}^{k+1}
S_-^{j j+k+1}) ,  \label{He}
\end{eqnarray}
where the effective interaction constants 
$\psi_j^{(k)}$ can be obtained from the recurrence relation 
\begin{equation}
\psi_j^{(k+1)} = \alpha_{j+k}^{(1)}\psi_j^{(k)}-
\alpha_j^{(1)} \psi_{j+1}^{(k)}, 
\quad .  \label{psi}
\end{equation}
with the initial term $\psi_j^{(1)} = g_j$. It is easy to see 
that $\psi_j^{(k)} \sim \alpha_{j}^{(k-1)}$ and thus 
$\psi_j^{(k+1)}\ll \psi_j^{(k)}$. 

The term $h_{{\rm diag}}$ contains only diagonal terms 
in the atomic operators and depends on the integral of motion 
$\hat{N}$ (or, which is equivalent, depends only on the 
photon-number operator $a^\dagger a$). This operator 
$h_{{\rm diag}}$ appears naturally represented as an 
expansion on the small parameter $\alpha_j^{(1)}$ whose
first term has the following form 
\begin{equation}
h_{{\rm diag}}= \frac{1}{2}
\sum_{j=1}^{N-1}g_j \alpha_j^{(1)}
[S_z^{j j+1} ( 2a^\dagger a+1) +
\{ S_+^{j j+1}, S_-^{j j+1} \}_+ ] + \mathcal{O}(\alpha^3). 
\end{equation}
The essential point is that this diagonal operator cannot be 
removed from the effective Hamiltonian (\ref{He}). 
On the contrary, the operator $h_{{\rm nondiag}}$ contains 
only nondiagonal terms that can be eliminated by 
rotations of the type (\ref{Tp}). In this respect, let us note 
that the price we must pay for eliminating one-photon transitions 
is the generation of all possible (in this atomic configuration) 
$k$-photon transitions ($k=2,...,N-1$). The possibility of
removing some of the terms in the sum appeared in (\ref{He}) 
strongly depends on the resonance conditions. In the next
section we shall examine this delicate point by means of 
some examples.

\section{Examples}

\subsection{Two-photon resonance.}

Let us consider the interaction of a collection of $A$ identical 
three-level atoms ($N=3$) in a cascade configuration with 
a single-mode quantum field of frequency $\omega _{f}$ and 
suppose that all the atomic transitions are far from the 
one-photon resonance with the field. Then, these transitions 
can be eliminated using the transformation (\ref{T1}). Taking 
into account the two-photon resonance condition $\Delta_3=0$ 
(i.e.; $\omega_3 - \omega_1 =2 \omega_f $) and keeping 
terms up to order $\alpha_j^{(1)}$, we obtain from~(\ref{He}) 
an effective Hamiltonian for two-photon  transitions 
(the nondiagonal part, $h_{{\rm nondiag}}$, can be eliminated 
by an appropriate transformation and no contributions 
of order $\alpha _{j}^{(1)}$ appear) 
\begin{equation}
H_{{\rm eff}}^{(1)} = h_0 + h_{{{\rm diag}}}
+ \frac{1}{2}
\psi _{1}^{(2)} (a^2 S_+^{13} + {a^\dagger}^ 2 S_-^{13} ) ,
\end{equation}
where 
\begin{equation}
h_{{{\rm diag}}}= \frac{1}{2} \sum_{j=1}^2 
g_j  \alpha_j^{(1)} [ S_z^{j j+1}
( 2a^\dagger  a + 1) +\{S_+^{j j+1}, S_-^{j j+1}\}_+ ] ,
\end{equation}
and the interaction constants, from (\ref{alfa}) and (\ref{psi}),
result to be 
\begin{eqnarray}
\label{ic1}
& \alpha _{1}^{(1)} = \frac{g_1}{\Delta_2}, \qquad 
\alpha_2^{(1)} = - \frac{g_2}{\Delta_2} , & \nonumber \\ 
&  &  \nonumber   \\
& \psi_1^{(2)} = \alpha_2^{(1)} g_1 - 
\alpha_1^{(1)} g_2 = - 2 \frac{g_1 g_2}{\Delta_2} . & 
\end{eqnarray}
Taking into account that $S^{11}+S^{22}+S^{33}=A$ and imposing
the absence of initial population in level $2$ (note that the term 
$h_0$ disappears in this case), we obtain the effective two-photon 
Hamiltonian including the dynamical Stark shift~\cite{11,12,13} 
\begin{eqnarray}
\label{H2f}
H_{{\rm eff}} & = & \frac{g_1 g_2}{\Delta_2 }
(a^2 S_{+}^{13} + {a^\dagger}^2 S_-^{13}) \nonumber  \\
&+& 
( S_z^{13} +A/2) 
\left [ \left( \frac{g_2^2 - g_1^2}{\Delta_2} \right ) 
a^\dagger a  + \frac{g_2^2}{\Delta_2} \right] +
A \frac{g_1^2}{\Delta_2} a^\dagger a .  
\end{eqnarray}

\subsection{Three-photon resonance}

Let us consider the interaction of a collection of $A$ identical 
four-level atoms ($N=4$) in a cascade configuration with a 
single-mode quantum field of frequency $\omega_f$ and 
suppose that there are no one- and two-photon resonances 
with the field.

After eliminating one-photon transitions the transformed 
Hamiltonian (\ref{He}) (written up to the corresponding order) 
has the form 
\begin{eqnarray}
\label{H3}
H_{{\rm eff}}^{(1)} & = & h_0 + h_{{\rm diag}} 
+ h_{{\rm nondiag}} \nonumber \\
& + & \frac{1}{2} 
\sum_{j=1}^2 \psi_j^{(2)} (a^2 S_+^{j j+2} + 
{a^\dagger}^2 S_-^{j j+2} ) +   
\frac{1}{3} \psi_1^{(3)}
(a^3 S_+^{14} + {a^\dagger}^3 S_-^{14} ) ,  
\end{eqnarray}
where 
\begin{equation}
 h_{{\rm diag}} = \frac{1}{2}\sum_{j=1}^3
g_j \alpha_j^{(1)} [S_z^{j j+1}( 2 a^\dagger a + 1)
+ \{S_+^{j j+1}, S_-^{j j+1}\}_+ ] + \mathcal{O}(\alpha^3), 
\end{equation}
and the essential part of the nondiagonal operator $h_{{\rm nondiag}}$
is 
\begin{equation}
h_{{\rm nondiag}} = \frac{1}{2} \sum_{i,j=1}^3 
\alpha_i^{(1)} g_j (S_+^{i i+1} S_-^{j j+1} +
S_+^{j j+1} S_-^{i i+1})  + \mathcal{O}(\alpha^2) . 
\end{equation}
The interaction constants are defined according to (\ref{alfa}) 
and (\ref {psi}): 
\begin{eqnarray}
& \alpha_1^{(1)} = \frac{g_1}{\Delta_2}, \qquad
\alpha_2^{(1)} = \frac{g_2}{\Delta_3 - \Delta_2}, \qquad
\alpha_3^{(1)} = - \frac{g_3}{\Delta_3}, \nonumber \\
& \psi_1^{(2)} = g_1 g_2 \frac{2\Delta_2 - \Delta_3}
{\Delta_2 ( \Delta_3 - \Delta_2 )} , \qquad
\psi_2^{(2)} = g_2 g_3 \frac{2 \Delta_3 - \Delta_2}
{\Delta_3 ( \Delta_2 - \Delta_3 )}, & \nonumber \\
& \psi_1^{(3)} = \frac{3g_1 g_2 g_3}{\Delta_3 \Delta_2}  , &
\end{eqnarray}
where, in addition, the resonance condition $\Delta_4=0$ (i.e.,
three-photon resonance $\omega_4 - \omega_1 = 
3\omega_f $) has been imposed.

According to the general scheme, the second term 
(representing two-photon transitions) in (\ref{H3}) can 
be removed by the unitary transformation  
\begin{equation}
U_2^{(1)}= \exp [T_{2}^{(1)}] ,
\end{equation}
with 
\begin{equation}
T_2^{(1)} = \frac{1}{2} 
\sum_{j=1}^2 \alpha_j ^{(2)}
(a^2 S_+^{j j+2} - {a^\dagger}^2 S_-^{j j+2}), 
\end{equation}
where now
\begin{equation}
\alpha_j^{(2)} = \frac{\psi_j^{2}}{\Delta_{j+2} - \Delta_j} 
\end{equation}
are also assumed to be small parameters; i.e., there are no resonant
two-photon transitions ($\Delta_{j+2} - \Delta_j = 
\omega_{j+2} - \omega_j - 2 \omega_f \gg \psi_j^{(2)}$) 
and thus $\alpha_j^{(2)}\ll \alpha_j^{(1)}$. 

The nondiagonal term $h_{{\rm nondiag}}$ can be also eliminated 
(in the same approximation) with the following transformation 
\begin{equation}
U_2^{(2)}= \exp [T_{2}^{(2)}] , 
\end{equation}
where 
\begin{equation}
T_2^{(2)} = \frac{1}{2} \sum_{i,j=1}^3
\beta_{ij} (S_+^{i i+1} S_-^{j j+1} + 
S_+^{j j+1} S_-^{i i+1}), 
\end{equation}
and 
\begin{equation}
\beta_{ij} = \frac{\alpha_i^{(1)} g_j}{\Delta_{i+1} - \Delta_i +
\Delta_j - \Delta_{j+1}}. 
\end{equation}
Taking into account that $S^{11}+S^{22}+S^{33}+S^{44}=A$ and
imposing the condition of the absence of initial population in levels 2
and 3, we finally obtain the effective Hamiltonian describing 
three-photon resonant transitions
\begin{eqnarray}
\label{H3f}
H_{{\rm eff}} &=&\frac{1}{3} \psi_1^{(3)} 
(a^3 S_+^{14} + {a^\dagger}^3 S_-^{14} ) \nonumber \\
& - & \frac{1}{2} [ \alpha_1^{(1)}g_1 
(a^\dagger a - 1/2 ) S^{11} - 
\alpha_3^{(1)} g_3 (a^\dagger a + 3/2 ) S^{44} ] \nonumber  \\
&- & \frac{1}{2} \beta_{31} S^{11} S^{44}\nonumber  \\
& - & \frac{1}{4} 
[ \alpha_1^{(2)} \psi_1^{(2)} a^\dagger a (a^\dagger a - 1) S^{11}
- \alpha_2^{(2)} \psi_2^{(2)} (a^\dagger a + 1) 
(a^\dagger a + 2) S^{44} ] .
\end{eqnarray}
One must note that the two last terms are of order 
$1/\Delta^3$ and, thus, can be omitted in the effective Hamiltonian, 
because the interaction term generating transitions between levels 1 
and 4  is of order $1/\Delta^2$. Then, the effective Hamiltonian takes 
the form 
\begin{eqnarray}
H_{{\rm eff}} & = & 
\frac{3g_1 g_2 g_3}{\Delta_2 \Delta_3} 
(a^3 S_+^{14} + {a^\dagger}^3 S_-^{14} ) \nonumber \\
& - & \frac{1}{2} \left[ 
\frac{g_1^2}{\Delta_2} 
( a^\dagger a - 1/2 )(A/2 - S_z^{14} ) + 
\frac{g_3^2}{\Delta_3} 
(a^\dagger a  + 3/2 ) 
(A/2 + S_z^{14})
\right ] ,  \label{H3f1}
\end{eqnarray}
where $S_z^{14} = (S^{44}-S^{11})/2$.

As in the two-photon case, the three-photon Hamiltonian (\ref{H3f1}) 
contains the dynamical Stark shift (with the corresponding
spontaneous contribution~\cite{11}). Nevertheless, the essential difference
between these two cases lies in the fact that in the two-photon case the
interaction term and the Stark-shift term are of the same order of
magnitude, while in the three-photon case the interaction term is one order
of magnitude lesser than the Stark-shift term. This  lead to 
essential differences in the evolution of some observables. In the 
same way, $k$-photon effective interaction Hamiltonian can be obtained.

\end{document}